\documentclass[twocolumn,9pt,numbers,round,sort&compress]{extarticle}
\pdfoutput=1

\usepackage{arxiv_style}

\usepackage[utf8]{inputenc} 
\usepackage[T1]{fontenc}    
\usepackage{hyperref}       
\usepackage{url}            
\usepackage{booktabs}       
\usepackage{amsfonts}       
\usepackage{amsmath}       

\usepackage{nicefrac}       
\usepackage{microtype}      

\usepackage[T1]{fontenc}
\usepackage[utf8]{inputenc}
\usepackage{authblk}

\usepackage{graphicx}
\usepackage{float}
\usepackage{authblk}
\usepackage[font=small,bf]{caption}

\title{Spatio--temporal coupling of attosecond pulses}

\author[1*]{Hampus Wikmark}
\author[1*]{Chen Guo}
\author[1]{Jan Vogelsang}
\author[2]{Peter W. Smorenburg}
\author[1]{H\'el\`ene Coudert-Alteirac}
\author[1]{Jan Lahl}
\author[1]{Jasper Peschel}
\author[1]{Piotr Rudawski}
\author[1]{Hugo Dacasa}
\author[1\textdagger] {Stefanos Carlstr\"om}
\author[1]{Sylvain Maclot}
\author[3]{Mette B. Gaarde}
\author[1]{Per Johnsson}
\author[1]{Cord L. Arnold}
\author[1\textdaggerdbl]{Anne L'Huillier}

\affil[1]{Physics Department, Lund University, SE-221 00, Lund, Sweden}
\affil[2]{ASML Netherlands B.V.,
De Run  6501,
5504 DR, Veldhoven,
The Netherlands}
\affil[3]{Department of Physics and Astronomy, Louisiana State University, Baton Rouge, Louisiana 70803-4001, USA}
\affil[*]{H.W. and C.G. contributed equally to this work.}
\affil[ \textdagger]{Current address: Max Born Institute, Max Born Strasse 2a, 12489 Berlin Adlershof, Berlin, Germany.}
\affil[ \textdaggerdbl]{To whom correspondence should be addressed. E-mail: anne.lhuillier@fysik.lth.se}

\begin{document}

\twocolumn[
  \begin{@twocolumnfalse}
    \maketitle
\begin{abstract}
The shortest light pulses produced to date are of the order of a few tens of attoseconds, with central frequencies in the extreme ultraviolet range and bandwidths exceeding tens of eV. They are often produced as a train of pulses separated by half the driving laser period, leading in the frequency domain to a spectrum of high, odd-order harmonics.
As light pulses become shorter and more spectrally wide, the widely-used approximation consisting in writing the optical waveform as a product of temporal and spatial amplitudes does not apply anymore.
Here, we investigate the interplay of temporal and spatial properties of attosecond pulses. We show that the divergence and focus position of the generated harmonics often strongly depend on their frequency, leading to strong chromatic aberrations of the broadband attosecond pulses. Our argumentation uses a simple analytical model based on Gaussian optics, numerical propagation calculations and experimental harmonic divergence measurements. This effect needs to be considered for future applications requiring high quality focusing while retaining the broadband/ultrashort characteristics of the radiation.    
\end{abstract}
\hrule \vspace{5pt}
  \end{@twocolumnfalse}
]


Electromagnetic waves are usually mathematically described by a product of purely spatial and purely temporal terms. This approximation often fails for broadband femtosecond laser pulses [see \cite{AkturkJO2010} and references therein] and spatio-temporal couplings need to be considered. Spatio--temporal couplings for visible or infrared light may be introduced by refractive and dispersive elements, such as lenses, gratings or prisms. The noncollinear amplification in optical parametric crystals may also potentially lead to spatio--temporal couplings, and it is important to develop characterization methods to measure and reduce their effects \cite{MirandaOL2014,ParienteNP2016,HarthJO2017}. In some cases, these couplings may be advantageously used, as, for example, demonstrated by Vincenti and Qu\'er\'e for the so--called ``lighthouse'' effect \cite{VincentiPRL2012,KimNP2013,LouisyO2015}.

The shortest light pulses, generated by high-order harmonic generation (HHG) in gases, are in the extreme ultraviolet (XUV)/ soft X-ray region and in the range of 100 as \cite{SansoneScience2006,GoulielmakisScience2008,LiNC2017,
GaumnitzOE2017}, with bandwidths of a few tens or even hundreds of eVs \cite{PopmintchevScience2012,CousinPRX2017}. These pulses are generated in a three-step process which was proposed at the beginning of the 1990's \cite{SchaferPRL1993,CorkumPRL1993}. When an atom is exposed to a strong laser field, an electron in the ground state can tunnel through the atomic potential bent by the laser field, propagate in the continuum, and recombine back to the ground state when (and if) returning close to the ionic core. In this process, an XUV photon is emitted, with energy equal to the ionization energy plus the electron kinetic energy at return. Two main families of trajectories leading to the same photon energy can be identified. They are characterized by the ``short'' or ``long'' time of travel of the electron in the continuum \cite{LewensteinPRA1995,BelliniPRL1998}. Interferences of attosecond pulses emitted at each laser half-cycle leads to a spectrum of odd-order harmonics.  

The investigation of spatio--temporal coupling of attosecond pulses requires measurements of their spatial properties, as a function of time or, equivalently, frequency. Wavefronts of high-order harmonics have been measured by several groups, using different techniques such as Spectral Wavefront Optical Reconstruction by Diffraction (SWORD) \cite{FrumkerOL2009, LloydSR2016, JohnsonScA2018}, lateral shearing interferometry \cite{AustinOL2011}, point-diffraction interferometry \cite{LeeOL2003} and Hartmann diffraction masks \cite{ValentinJOSA2008,FreisemOE2018}. In particular, Frumker et al. \cite{FrumkerOE2012} pointed out that the variation of wavefront and intensity profile with harmonic order leads to spatio--temporal coupling of the attosecond pulses, with temporal properties depending on where they are measured.

The spatial and spectral properties of high-order harmonics strongly depend on the geometry of the interaction, and in particular on whether the gas medium in which the harmonics are generated, is located before or after the focus of the driving laser beam \cite{SalieresPRL1995}. The asymmetry between ``before'' and ``after'' can be traced back to the phase of the emitted radiation, which is equal to that of the incident laser field multiplied by the process order, as in any upconversion process, plus the dipole phase which is accumulated during the generation and mostly originates from electron propagation in the continuum. While the former is usually antisymmetric relative to the laser focus, the latter depends on the laser intensity and is therefore symmetric \cite{BalcouPRA1997,AustinOL2011}. The total phase and thus the divergence properties are different before and after the laser focus, leading to a strong dependence of the spatio--temporal properties of the harmonic radiation on the generation conditions. In some conditions, harmonics can be emitted with a flat wavefront \cite{AustinOL2011} or even as a converging beam \cite{Quintard2017,Quintard2018}. Another phenomenon leading to an asymmetry of HHG with respect to the generation conditions is ionization-induced reshaping of the fundamental field, which depends on whether the field is converging or diverging when entering the gas medium \cite{MiyazakiPRA1995,TamakiPRL1999,LaiOE2011,JohnsonScA2018}.
 
\begin{figure}[h]
\centering
\mbox{\includegraphics[width=\linewidth]{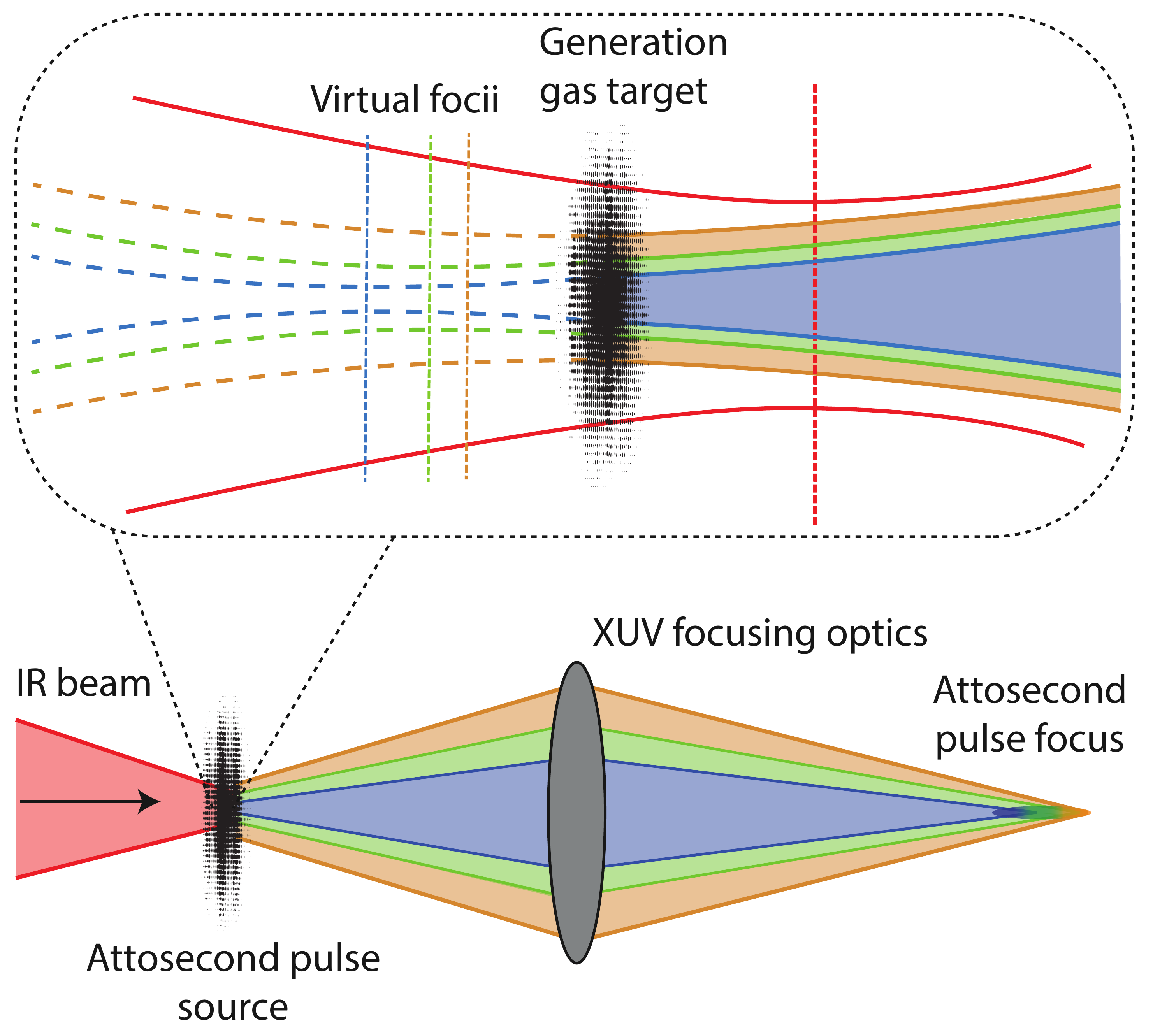}}
\caption{Illustration of spatio-temporal coupling for an attosecond pulse: different frequencies (indicated here by three colors, blue, green, orange), generated with varying wavefront curvatures and different divergences, as indicated in the inset, will be refocused by XUV optics (here represented as a lens) at different positions, leading to strong chromatic aberrations and an extended focus, both transversally and longitudinally. The fundamental driving field is indicated by the red line. }
\label{fig:coupling}
\end{figure}

In the present work, we show that the frequency components of attosecond pulses generated by HHG in gases have different divergence properties, which depend on the geometry of the interaction and in particular on where the generating medium is located relative to the laser focus. In some conditions, the position of the focus and divergence strongly vary with frequency, leading to chromatic aberrations, as sketched in the inset in Fig.~\ref{fig:coupling}, similar to the effect that a chromatic lens has on broadband radiation\cite{BorOL1989,BorOC1992}.
Any imaging optical component [see Fig.~\ref{fig:coupling}] will focus the frequency components of the attosecond pulses at different locations, resulting in spatio-temporal couplings. Depending on the position where the pulses are characterized or utilized, they will have different central frequencies, pulse durations and spatial widths. We develop an analytical model based on an analytical expression for the dipole phase \cite{GuoJPBAMOP2018} combined with traditional Gaussian optics to predict the radius of curvature, position of focus and divergence of the two trajectory contributions to HHG. This model is validated using numerical simulations of HHG \cite{LhuillierPRA1992} for both thin and thick generating media. We also present experimental measurements of the harmonic divergence as a function of position of generation relative to the laser focus. Finally, we discuss the implications of our results for the focusing of broadband attosecond pulses. 

\section*{Analytical expression of the dipole phase}
\label{sec:phase}

The single atom response of HHG is well described by an approximate solution of the time-dependent Schr\"odinger equation for an atom in a strong laser field, called the Strong-Field Approximation (SFA) \cite{LewensteinPRA1994}. This theory leads to a simple analytical expression of the dipole phase, equal to $\alpha I$, where $\alpha$ depends on the harmonic order and on the trajectory contributing to HHG \cite{SalieresPRL1995,LewensteinPRA1995,VarjuJMO2005,CarlstromNJP2016} and where $I$ is the laser intensity. This expression has been used in numerous investigations of the harmonic properties \cite{BelliniPRL1998,ZairPRL2008,CarlstromNJP2016,Quintard2018}.
Here, we utilize a more general analytical expression for the phase \cite{GuoJPBAMOP2018}, based on the semi-classical description of attosecond pulse generation \cite{SchaferPRL1993,CorkumPRL1993}.

\begin{figure}[htbp] 
\centering
\mbox{\includegraphics[width=0.9\linewidth]{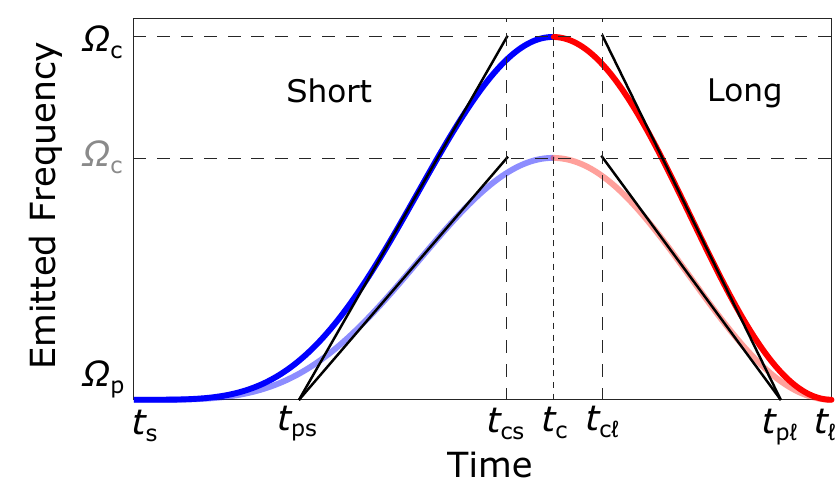}}
\caption{Emitted XUV frequency as a function of return time for two laser intensities, corresponding to the blue/red and light blue/red solid curves. The blue curves describes the short trajectories, while the red lines refer to the long trajectory. $t_{\mathrm{s},\ell}$ are the return times for the short and long electron trajectories leading to the threshold frequency $\Omega_\mathrm{p}$. $t_\mathrm{c}$ is the return time for the trajectory leading to the cutoff frequency $\Omega_\mathrm{c}$. $t_{\mathrm{p}i}$ and $t_{\mathrm{c}i}$ ($i$=s,$\ell$) are return times obtained by approximating $\Omega(t)$ as piecewise straight lines. Values for these return times are indicated in Table~\ref{tab:parameters_x}.}
\label{fig:energy}
\end{figure}
 
\begin{table}[htbp]
\centering
\caption{Return times for the short and long trajectories relative to the zero of the electric field. For the last column, a laser wavelength of 800 nm is used.\\}
\begin{tabular}{llcc}
\hline
Units  & Brief description & (Cycle) & (fs) \\
$t_\mathrm{s}$ & short, threshold& $0$ & $0$ \\
$t_\mathrm{ps}$ & short, threshold, model & $0.18$ & $0.48$ \\
$t_\mathrm{cs}$  & short, cutoff, model & $0.40$ & $1.07$\\
$t_\mathrm{c}$ & cutoff & $0.45$ & $1.20$\\
$t_{\mathrm{c}\ell}$ & long, cutoff, model & $0.50$ & $1.35$\\
$t_{\mathrm{p}\ell}$ & long, threshold, model & $0.69$ & $1.85$\\
$t_\ell$ & long, threshold & $0.75$ & $2.00$ \\
\hline
\end{tabular}
  \label{tab:parameters_x}
\end{table}

In this approximation, the second step of the process is described by solving Newton's equation of motion for a free particle in the laser field. Figure \ref{fig:energy} shows the frequency ($\Omega$) of the emitted XUV radiation as a function of electron return time for two different fundamental field intensities, indicated by the bright or faint colors. The frequency varies from $\Omega_\mathrm{p}$,   corresponding to the ionization threshold ($\hbar\Omega_\mathrm{p}$=$I_\mathrm{p}$, $I_\mathrm{p}$ denoting the ionization energy and $\hbar$ the reduced Planck constant) to the cutoff frequency $\Omega_\mathrm{c}$ ($\hbar\Omega_\mathrm{c}= 3.17 U_\mathrm{p}+I_\mathrm{p}$). $U_\mathrm{p}$ denotes the ponderomotive energy, equal to
\begin{equation}
U_\mathrm{p}= \frac{\alpha_{\scalebox{.5}{\textsc{FS}}} \hbar I \lambda^2}{2 \pi c^2 m},
\end{equation}
where $\alpha_{\scalebox{.5}{\textsc{FS}}}$ is the fine structure constant, $m$ the electron mass, $c$ the speed of light and $\lambda$ the laser wavelength. 
The frequency variation can be approximated by piecewise straight lines as indicated by the black solid lines. After inversion from $\Omega(t)$ to $t(\Omega)$, for each straight line, we have 
\begin{equation}
t_i(\Omega) = t_{\mathrm{p}i}+ \frac{t_{\mathrm{c}i}-t_{\mathrm{p}i}}{\Omega_\mathrm{c}-\Omega_\mathrm{p}}(\Omega-\Omega_\mathrm{p}),
\end{equation}
where $i$=s,$\ell$ refers to the electron trajectory (short or long) and $t_{\mathrm{p}i}$, $t_{\mathrm{c}i}$ are defined as indicated by the dashed lines in Fig.~\ref{fig:energy}. The values of $t_{\mathrm{p}i}$ and $t_{\mathrm{c}i}$, in both laser cycles and femtoseconds (at $\lambda=800$~nm), are summarized in Table~ \ref{tab:parameters_x}. We also indicate the return times for the short and long electron trajectories leading to the threshold frequency ($t_\mathrm{s}$,$t_\ell$) and the return time for the trajectory leading to the cutoff frequency ($t_\mathrm{c}$). Neglecting the frequency dependence of the time for tunneling and recombination, $t_i(\Omega)$ can be interpreted as the group delay of the emitted radiation. Its integral is the spectral phase 
\begin{equation}
\Phi_i(\Omega)=\Phi_i(\Omega_\mathrm{p})+t_{\mathrm{p}i}(\Omega-\Omega_\mathrm{p})+\frac{t_{\mathrm{c}i}-t_{\mathrm{p}i}}{\Omega_\mathrm{c}-\Omega_\mathrm{p}}\frac{(\Omega-\Omega_\mathrm{p})^2}{2}.
\label{eq:phase}
\end{equation}

As shown in Fig.~\ref{fig:energy}, the return times $t_{\mathrm{p}i}$, $t_{\mathrm{c}i}$, and therefore the second term in Eq.~[\ref{eq:phase}] do not depend on laser intensity. Using $\Omega_\mathrm{c}-\Omega_\mathrm{p}=3.17U_\mathrm{p}/\hbar$, the coefficient in the third term can be written as
\begin{equation}
\frac{t_{\mathrm{c}}-t_{\mathrm{p}i}}{\Omega_\mathrm{c}-\Omega_\mathrm{p}}= \frac{2\gamma_i}{I},
\end{equation}
where 
\begin{equation}
\gamma_i=\frac{(t_{\mathrm{c}i}-t_{\mathrm{p}i})\pi c^2 m}{3.17 \alpha_{\scalebox{.5}{\textsc{FS}}}\lambda^2},
\end{equation}
In this  classical calculation, $\Phi_i(\Omega_\mathrm{p})$ is equal to zero for the short trajectory, while it is proportional to the laser intensity for the long: $\Phi_{\ell}(\Omega_\mathrm{p})=\alpha_{\ell} I$. The value of $\alpha_{\ell}$ can be obtained numerically within the classical approach used in this work \cite{Guo2018}, and is found to be close to that given within the SFA, equal to $4\pi^2\alpha_{\scalebox{.5}{\textsc{FS}}}/m\omega^3$ \cite{LewensteinPRA1995,CarlstromNJP2016}. 
Table \ref{tab:parameters} indicates the parameters needed to describe $\Phi_i(\Omega)$ for 800 nm radiation.  

\begin{table}[htbp]
\centering
\caption{\bf Parameters for the short and long trajectories at 800 nm}
\begin{tabular}{ccc}
\hline
$\gamma_\mathrm{s}$ & $1.03\times 10^{-18} \,\mathrm{s}^2\mathrm{W cm}^{-2}$\\
$\gamma_\ell$ & $-0.874\times 10^{-18}\,\mathrm{s}^2\mathrm{W cm}^{-2}$\\
$\alpha_\mathrm{s}$ & $0$ \\ 
$\alpha_\ell$ & $-2.38\times 10^{-13} \,\mathrm{W}^{-1}\mathrm{cm}^{2}$\\ 
\hline
\end{tabular}
  \label{tab:parameters}
\end{table}

The dipole phase can be approximated for the two families of trajectories by the expansion:
\begin{equation} 
\Phi_i(\Omega)=\alpha_iI+t_{\mathrm{p}i}(\Omega-\Omega_\mathrm{p})+ \frac{\gamma_i}{I}(\Omega-\Omega_\mathrm{p})^2.
\label{eq:phasef}
\end{equation}
The present expression gives very similar results to e.g. the numerical results presented in \cite{VarjuJMO2005}, obtained by solving saddle point equations within the SFA, with the advantage of being analytical. 

\section*{Wavefront and spatial width of XUV radiation}

 We now use this analytical expression for the dipole phase together with traditional Gaussian optics to predict the radius of curvature, position of focus and divergence of the two trajectory contributions to HHG. A similar derivation has been  proposed, independently, by Quintard et al. \cite{Quintard2017,Quintard2018} with, however, a different analytical formulation of the dipole phase. We neglect the influence of propagation, considering an infinitely thin homogeneous gas medium, and assume that the fundamental field is Gaussian, with intensity $I(r,z)$, width $w(z)$, radius of curvature $R(z)$ and peak intensity $I_0$, $z$ denoting the coordinate along the propagation axis and $r$ the radial coordinate. The focus position is $z=0$ and the waist $w_0=w(0)$. Considering only the contribution of one trajectory~$i$, the phase of the $q^{\mathrm{th}}$ harmonic field can be approximated by 
\begin{equation}
 \Phi_q(r,z)= q\phi(r,z)+ \Phi_i(r,z). 
\end{equation}
The phase of the fundamental Gaussian beam is $\phi(r,z)=kz-\zeta(z)+kr^2/2R(z)$, where $k$ is the wavevector equal to $\omega/c$ and $\zeta(z)$ the Gouy phase \cite{SalehPhotonics2007}. This article is mainly concerned with the third term, giving the curvature of the beam. The dipole phase $\Phi_i(r,z)$ is given by Eq.~(\ref{eq:phasef}), for $I=I(r,z)$ and $\Omega=q\omega$, $\omega$ being the laser frequency. Omitting the second term in Eq.~(\ref{eq:phasef}), which does not depend on intensity and therefore on space, $\Phi_i(r,z)$  can be expressed as
\begin{equation}
\Phi_i(r,z)= \frac{\alpha_iI_0 w_0^2}{w^2(z)}e^{-\frac{2r^2}{w^2(z)}}+\frac{\gamma_i (\Omega-\Omega_p)^2w^2(z)}{I_0 w_0^2}e^{\frac{2r^2}{w^2(z)}}.
\label{eq:phiq}
\end{equation}

\begin{figure}[htbp] 
\centering \vspace{-5pt}
\mbox{\includegraphics[width=0.9\linewidth]{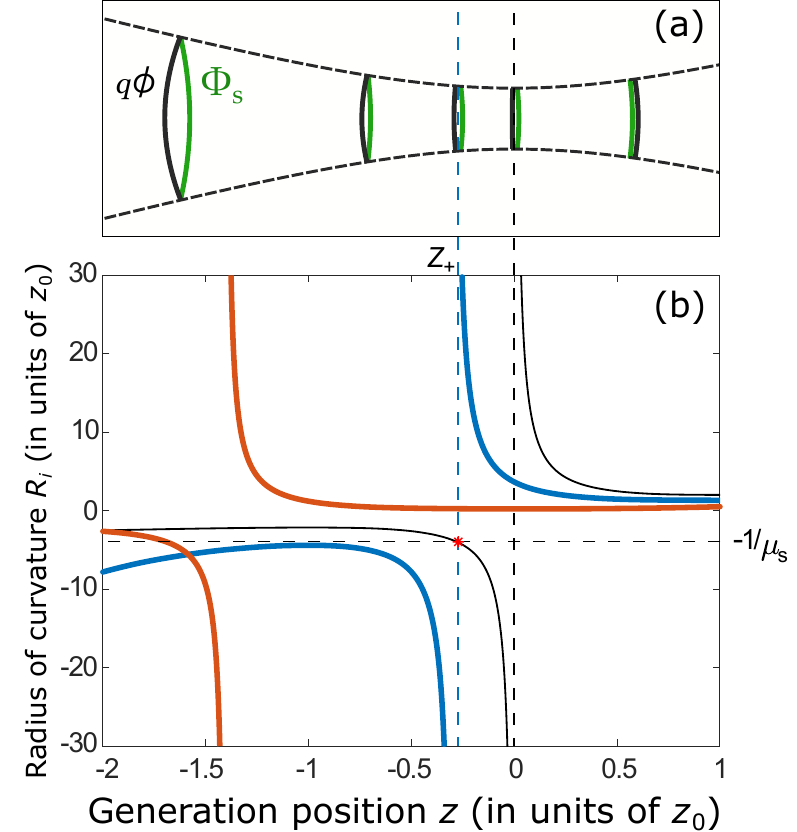}}
\caption{(a) Representation of different contributions to the harmonic wavefront, due the fundamental (black) and due to the dipole phase for the short trajectory (green) at different generation positions ($z$). The fundamental beam profile variation is indicated by the black dashed line. (b) Radius of curvature of the 23\textsuperscript{rd} harmonic as a function of generation position. The laser wavelength is 800 nm and the peak intensity at focus is $3 \times 10^{14}$ W cm$^{-2}$. The blue (red) solid line is obtained for the short (long) trajectory. The thin solid line shows the radius of curvature of the fundamental. At the position $Z_+$, where $R(z)=-z_0/\mu_\mathrm{s}$, $R_\mathrm{s}/z_0$ diverges. As can be seen in a) this is when the two phase contributions cancel out.}
\label{fig:radius}
\end{figure}

We use a Taylor expansion close to the center of the beam to approximate $\Phi_i(r,z)$ [Eq.~(\ref{eq:phiq})]. To determine the harmonic wavefront, we only keep the terms proportional to $r^2$ in Eq.~(\ref{eq:phiq}), to which we add the $r^2$-dependent contribution from the fundamental, equal to $qkr^2/2R(z)$. The resulting  $r^2$-dependent contribution to the phase of the harmonic field can be written as $qkr^2/2R_i$, with
\begin{equation}
\frac{1}{R_i}=\frac{1}{R(z)}-\frac{4\alpha_i I_0 w_0^2 c}{ w^4(z)\Omega}+\frac{4\gamma_i(\Omega-\Omega_p)^2c}{I_0 w_0^2\Omega}.
\label{eq:radius}
\end{equation}
For simplicity of the notations, we omit to explicitly indicate the $z$ dependence of $R_i$.  The curvature of the harmonic field is equal to that of the fundamental (first term) plus that induced by the dipole phase. The second term is only present for the long trajectory. This equation outlines the dependence of the XUV radiation wavefront on frequency ($\Omega$), electron trajectory (i), intensity at focus ($I_0$) and generation position ($z$). Eq.~(\ref{eq:radius}) is illustrated in Fig.~\ref{fig:radius}(a), representing the wavefronts induced to the harmonic by the fundamental (black) and due to the dipole phase for the short trajectory (green) as a function of the generation position. The fundamental wavefront changes from convergent to divergent through the focus, while that induced by the dipole phase is always divergent and independent of the generation position ($z$). 

Using the reduced coordinate $Z=z/z_0$, where $z_0=\pi w_0^2/\lambda$ is the fundamental Rayleigh length, Eq.~(\ref{eq:radius}) can be written as
\begin{equation}
\frac{z_0}{R_i}=\frac{1}{Z+1/Z}-\frac{\eta_i}{(1+Z^2)^2} +\mu_i,
\label{eq:redrad}
\end{equation}
 where $\eta_i=2\alpha_i I_0 /q$ and $\mu_i=2 \gamma_i \omega^2 (q-q_\mathrm{p})^2/qI_0$ are dimensionless quantities ($q_\mathrm{p}=\Omega_\mathrm{p}/\omega$). For the short trajectory, since $\alpha_\mathrm{s}=0$, the positions where the radius of curvature diverges, corresponding to a flat phase front, can be calculated analytically by solving a second-order equation in $Z$, 
\begin{equation}
Z^2+\frac{Z}{\mu_\mathrm{s}}+1=0.
\label{z2}
\end{equation}
For $\mu_\mathrm{s}\leq 0.5$, the solutions to this equation are real and the radius of curvature diverges at  
\begin{equation}
Z_\pm=-\frac{1}{2\mu_\mathrm{s}}\pm\sqrt{\frac{1}{4\mu^2_\mathrm{s}}-1}. 
\end{equation}

This discussion is illustrated graphically in  Fig.~\ref{fig:radius}(b) for the 23\textsuperscript{rd} harmonic of 800 nm radiation generated in argon, with $I_0=3 \times 10^{14}\mathrm{W}\,\mathrm{cm}^{-2}$. In these conditions, we have $\eta_\mathrm{s}=0$, $\mu_\mathrm{s}=0.253$, $\eta_\ell=-6.38$ and $\mu_\ell=-0.215$. Fig.~\ref{fig:radius}(b) presents the radius of curvature in reduced units $R_i/z_0$ for the short (blue) and long (red) trajectory contributions. Over the range shown in the figure, between $-2z_0$ and $z_0$,  $R_\mathrm{s}/z_0$, represented by the blue curve, diverges at $Z_+=-0.272$. The other solution of Eq.~(\ref{z2}) is $Z_-=-3.68$ which is outside the scale of the figure. For the long trajectory, the radius of curvature, represented by the red solid line, diverges at $Z \simeq -1.4$. This behavior is quite general for all harmonics, as discussed in the last section of this article. 

To estimate in a simple way the spatial width of the harmonic field at the generation position, we assume that its amplitude is proportional to the fundamental amplitude to a power $p$. This exponent is quite constant in the plateau region (typically of the order of 4) \cite{Quintard2018,DurfeePRL1999} and increases in the cutoff region. The harmonic width is then simply equal to $W=w(z)/\sqrt{p}$. (Here as well, we omit to write explicitly the $z$-dependence of $W$). 

\section*{Focus position and beam waist}

Knowing the beam radius of curvature and width at a given position $z$, it is a simple exercise within Gaussian optics to determine the position of the focus and the corresponding waist [see e.g. \cite{SalehPhotonics2007}]. The position of focus relative to the generation position $z$ is given by
\begin{equation}
z_i=-\frac{R_i}{1+(\lambda_q R_i /\pi W^2)^2},
\label{zi}
\end{equation}
with $\lambda_q=\lambda/q$. Using reduced coordinates relative to the fundamental Rayleigh length, Eq.~[\ref{zi}] can be written as
\begin{equation}
\frac{z_i}{z_0}=-\frac{R_i}{z_0}\left(1+\left[\frac{p R_i}{ q z_0(1+Z^2)}\right]^2\right)^{-1}.
\end{equation}
The corresponding waist at focus is 
\begin{equation}
w_i=\frac{W}{\sqrt{(1+(\pi W^2/\lambda_q R_i)^2}},
\end{equation}
or, relative to the fundamental waist,
\begin{equation}
\frac{w_i}{w_0}=\left(\frac{1+Z^2}{p}\right)^{\frac{1}{2}}\left(1+\left[\frac{qz_0 (1+Z^2)}{ pR_i}\right]^2\right)^{-\frac{1}{2}}.
\label{wi}
\end{equation}

\begin{figure}[h] 
\centering
\includegraphics[width=0.9\linewidth]{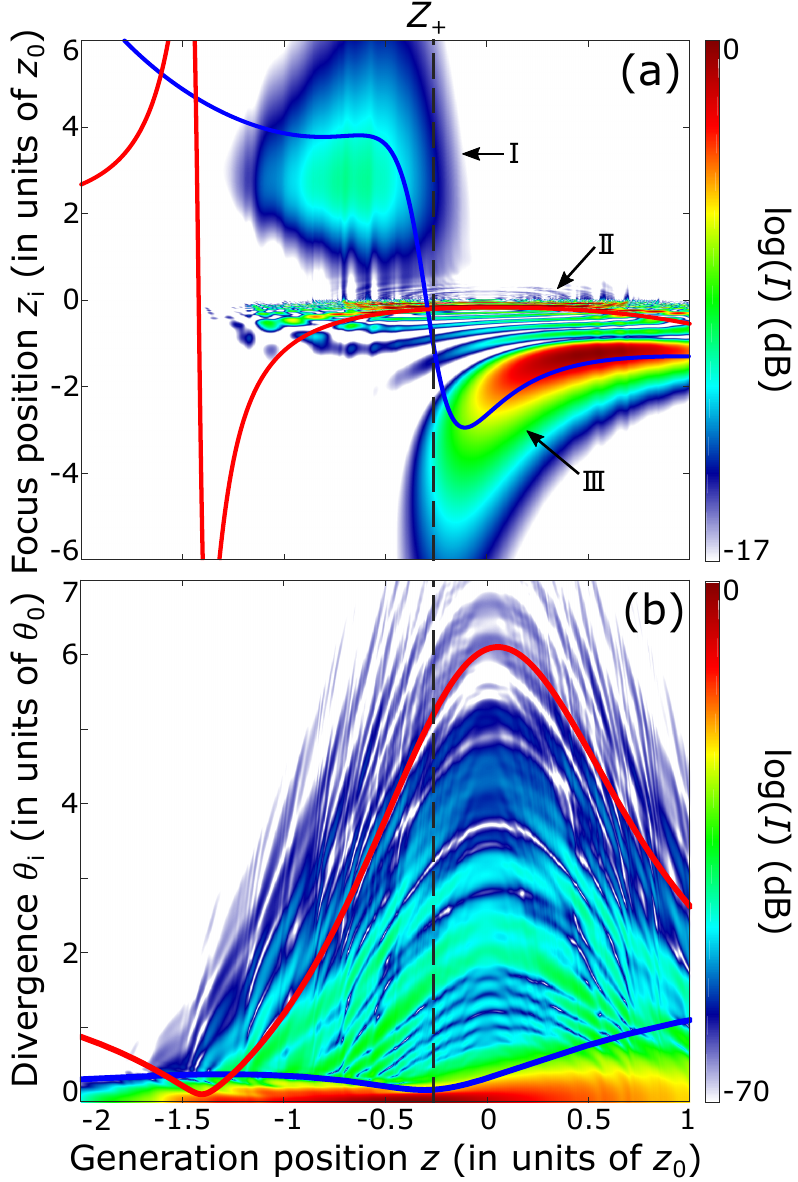}
\caption{Position of the focus of the 23\textsuperscript{rd} harmonic relative to the generation position (a) and far-field divergence (b) as a function of the generation position relative to the laser focus. The results for the short and long trajectory are indicated by the blue and red curves respectively. The dashed line corresponds to the position $Z_+$ where the radius of curvature for the short trajectory diverges. The color plots indicate results of a  calculation based on the solution of the TDSE, where HHG is assumed to occur in an infinitely thin plane. In (a), the on-axis intensity at a certain position along the propagation axis is plotted as a function of generation position on a logarithmic scale. Three different focal  regions, labeled I, II and III can be identified. In (b), the radial intensity calculated at a long distance from the generation position, and normalized to the fundamental radial intensity at the same distance is indicated.}
\label{fig:focus_tdse}
\end{figure}

Fig.~\ref{fig:focus_tdse} shows (a) the position of the harmonic focus ($z_i/z_0$) relative to that of the generation position ($z/z_0$), and (b) the normalized far-field divergence $\theta_i/\theta_0=w_0/w_i$ for the two trajectories, short (blue solid line) and long (red solid line). The color plots will be discussed in the next Section. The divergence of the fundamental $\theta_0$ is defined as $\lambda/\pi w_0$. Let us emphasize that the zero of the horizontal scale is the laser focus, while in (a), zero on the vertical scale means that the focus of the harmonic field coincides with the generation position. The focus position and divergence strongly vary with $z$, and quite differently for the two trajectories. In both cases, the focus position changes sign and the divergence goes through a minimum when the radius of curvature goes to infinity (see Fig.~\ref{fig:radius}).

For the short trajectory and $Z\leq Z_+$, the focus is real and it is located after the generation position ($z_i \geq 0$) along the propagation direction. The negative curvature of the convergent fundamental beam is larger in magnitude than the positive curvature induced by the dipole phase and the harmonics are generated as a convergent beam \cite{Quintard2018}. When $Z > Z_+$, the focus is virtual and located before the generation position. Two cases can be considered: When $0>Z>Z_+$, i.e. when the generation position is before the IR focus, the negative curvature of the fundamental beam is smaller in magnitude than the positive curvature induced by the dipole phase: The harmonics are generated as a divergent beam. When $Z \geq 0$, both curvatures are positive and the harmonics are generated as a divergent beam. The divergence is smallest in the region close to $Z_+$. 

The same reasoning applies for the long trajectory contribution, except that $Z_+$ is now replaced by $Z\approx -1.4$ (see Fig.~\ref{fig:radius}). In this case, in the region with enough intensity for HHG, i.e. $|Z| \leq 1.5$, corresponding to $I=9 \times 10^{13}$~W$\,$cm$^{-2}$, the harmonic focus is located just before the generation position and the divergence is much larger than that of the short trajectory contribution.

 At the positions where the radius of curvature diverges (indicated by the dashed line in Fig.~\ref{fig:focus_tdse} for the short trajectory), the harmonics are generated with a flat wavefront and with a large focus (low divergence). In contrast, harmonics generated far away from the divergence minima will inherit a curvature from the fundamental and the dipole contribution which corresponds to a significantly smaller beam waist in the real or virtual focus and thus in a significantly larger divergence. The variation of the divergence with generation position is due partly to the dipole phase contribution, but also to the mismatch between the harmonic order $q$ and the amplitude variation here described by a power law with exponent $p=4$ [See Eq.~(\ref{wi})].

\section*{Numerical calculations}

To validate the Gaussian model presented in this work, we performed calculations based on tabulated single-atom data obtained by solving the TDSE for a single active electron in argon. The time-dependent dipole response was calculated for $\simeq$ 5000 intensity points. This allows us, for each harmonic frequency, to precisely unwrap the amplitude and phase variation as a function of intensity, and thus to accurately describe the interferences of the trajectories. The complex electric field distribution at a given harmonic frequency is obtained by integrating in time the polarization induced by the fundamental field in an arbitrarily thin sheet of homogeneous argon gas. The field is then propagated to different positions relative to the generation position by calculating the diffraction integral in Fresnel approximation using Hankel transforms. The influence of ionization is not taken into account.  This procedure is repeated for different gas target positions relative to the laser focus. We use a fundamental wavelength of 800\,nm, a pulse duration of 45\,fs, a peak intensity of $3\times 10^{14}$W~cm$^{-2}$ and a fundamental waist size $w_0=350\,\mu$m. The corresponding Rayleigh length is equal to 0.48 m.

\begin{figure}[h] 
\centering
\includegraphics[width=0.9\linewidth]{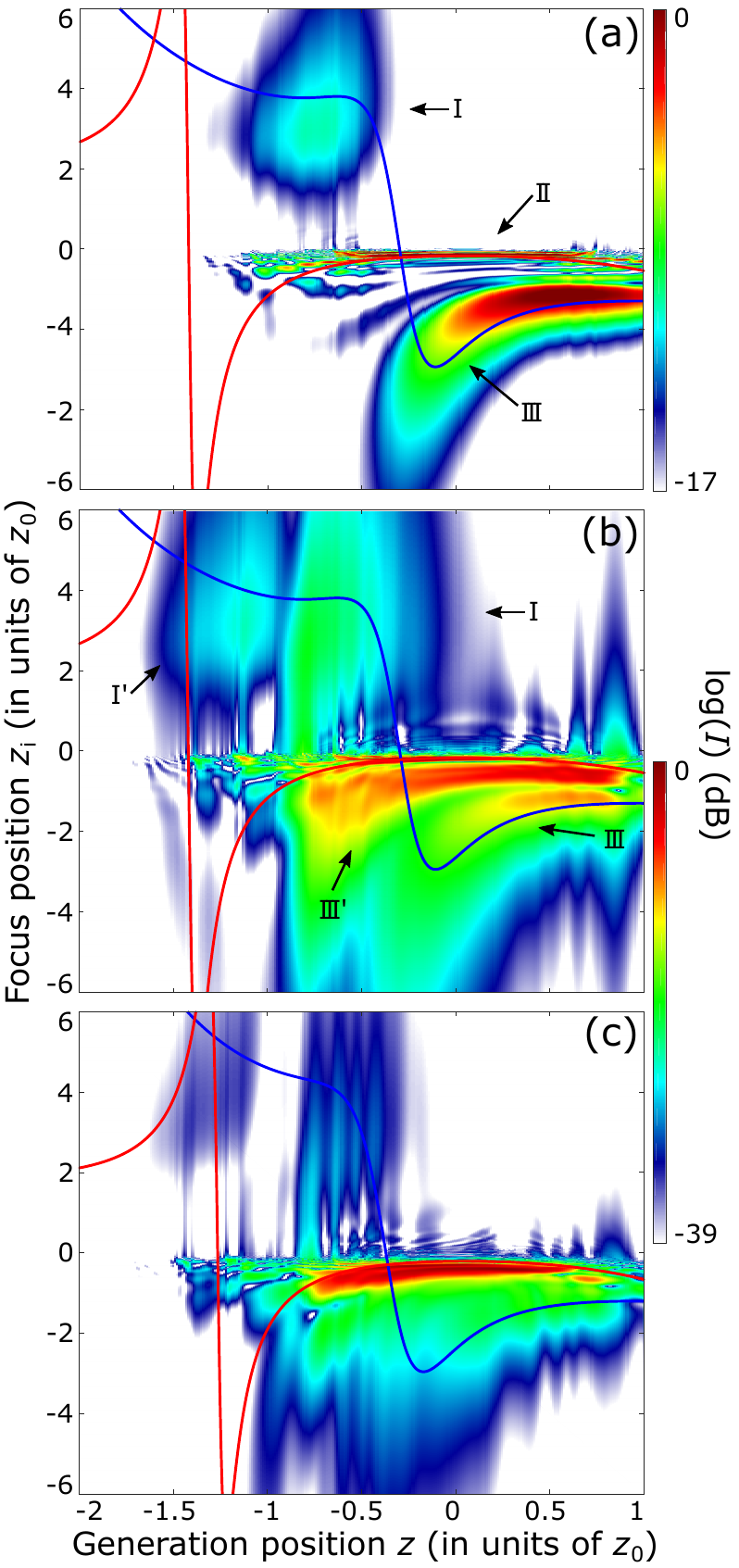}
\caption{Results of propagation calculations for a (a) 5.4 mm, (b) 30 mm and (c) 60 mm-long gas cell. The on-axis intensity at a certain position along the propagation axis axis is plotted as a function of generation position on a logarithmic scale. The results of the Gaussian model are indicated by the blue and red solid lines for the short and long trajectories and are identical to those of Fig.~\ref{fig:focus_tdse}(a).}
\label{fig:prop_tdse}
\end{figure}

Figure\,\ref{fig:focus_tdse} (a) presents a color plot of the 23$^\mathrm{rd}$ harmonic on-axis intensity for different generation positions (horizontal axis). The regions with the warmest colors (i.e. towards red) represent the focal regions. The small regions with high peak intensity (dark red, like that labeled II) correspond to the smallest focus. The agreement between the numerical predictions and those of the Gaussian model is striking. When $Z \leq Z_+$, the 23$^\textrm{rd}$ harmonic is focused after the generation position (region I). When $Z \geq Z_+$, two focal regions can be identified, a very thin one close to the generation position [region II], and a larger one at larger negative $z_i$ [region III]. The agreement with the results of the Gaussian model allows us to interpret the main contribution to these regions: short trajectory for I and III and long trajectory for II. The horizontal interference structures observed between I and II are a manifestation of quantum path interferences \cite{ZairPRL2008}. The harmonic radiation often exhibits two foci, due to the two trajectories. While the focus position for the long trajectory contribution remains close to (just before) the generation plane, the  focus position of the short trajectory contribution strongly depends on the generation position. The color plot in Fig.~\ref{fig:focus_tdse}(b) is the 23$^\textrm{rd}$ harmonic radial intensity at a distance of $50z_0$, as a function $r/50z_0\theta_0$ (vertical scale) for different generation positions. This distance is long enough to reach the far field region, so that the radial intensity is proportional to the far field divergence. As for the focus position, the comparison with the prediction of the Gaussian model allows us to distinguish the contribution of the two trajectories, with quite different divergence, especially for $|Z| \leq 1$. The red (blue) curves represent the $1/e^2$ divergence within the Gaussian model for the long (short) trajectories. The blue-green colored regions in (b) can be attributed to the long trajectory while the red-yellow-bright green regions to the short trajectory.

An important question is whether these results are still valid after propagation in a finite medium. We used the single atom data described previously as input in a propagation code based on the slowly-varying envelope and paraxial approximations \cite{LhuillierPRA1992}. We present in Fig.~\ref{fig:prop_tdse} results obtained for a 5.4 mm (a), 30 mm (b) and 60 mm (c)-long homogeneous medium, using a 2 mbar gas pressure. While Fig.~\ref{fig:prop_tdse}(a) compares very well with the results shown in Fig.~\ref{fig:focus_tdse}(a), as expected, Fig.~\ref{fig:prop_tdse}(b) and (c) shows clear effects of propagation, related to ionization-induced defocusing of the fundamental laser beam. In fact, two different phase matching regimes appear: one similar to what is present in absence of propagation, and which agrees well with the predictions of the Gaussian model [Compare regions I,III in Fig.~\ref{fig:prop_tdse}(a,b)] and a second one, which also follows a similar model but for a fundamental focus moved to the left [See regions I',III' in Fig.~\ref{fig:prop_tdse}(b)], as expected for a  fundamental beam that is defocused due to partial ionization of the medium \cite{MiyazakiPRA1995,TamakiPRL1999,LaiOE2011,JohnsonScA2018}. To examine in more details the effect of propagation goes beyond the scope of this article and will be discussed in future work.

\section*{Experimental divergence measurements}

Experiments were performed at the intense XUV beamline of the the Lund Laser Centre \cite{ManschwetusPRA2016,CoudertAS2017}, using a multi-terawatt 45-fs Titanium-Sapphire laser operating at 10~Hz repetition rate. The beam was (slightly) apertured to 27~mm and focused using a spherical mirror with focal length $f=8$~m. In addition, a deformable mirror was used in order to correct for the laser wavefront aberrations and adjust the focal length. The harmonics were generated in a 60~mm gas cell filled with argon by a pulsed valve. We measured the divergence of the emitted harmonics using a flat field XUV spectrometer with an entrance slit located approximately 6~m after the generation. For each harmonic, the width was estimated by fitting a Gaussian function onto the transverse (spatial) direction of the spectrometer. The IR focus was moved relative to the gas cell along the direction of propagation by changing the voltage of the actuators controlling the curvature of the deformable mirror.
 The limits of the scan were imposed by the decrease of the harmonic yield, which is slightly asymmetric relative to the laser focus \cite{SalieresPRL1995}. 

\begin{figure}[h!] 
\centering \vspace{-8pt}
\mbox{\includegraphics[width=0.9\linewidth]{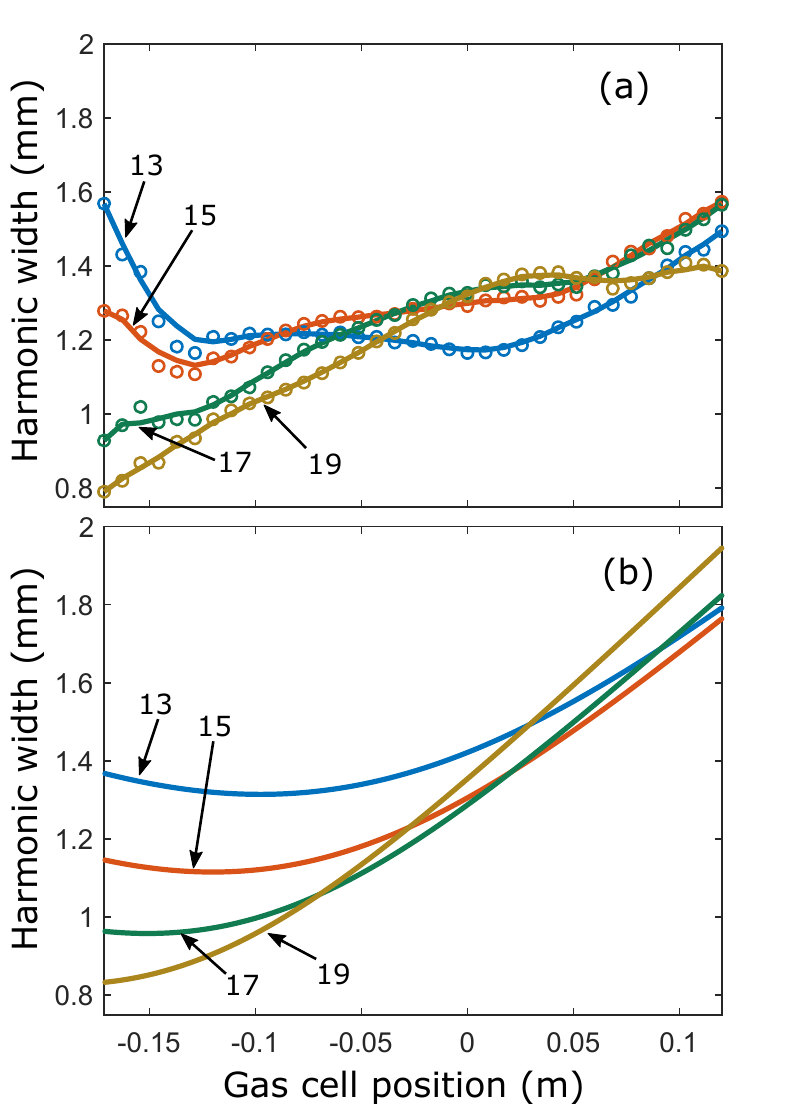}}
\caption{(a) Spatial widths of harmonics 13 to 19 generated in Ar and measured approximately 6 m after the generation as a function of the cell position. The solid lines are fit to the experimental data indicated by the circles. (b) Spatial widths of the same harmonics as a function of generation position, predicted by the Gaussian model. } \vspace{-5pt}
\label{fig:divergenceExperiment}
\end{figure}

The widths of the 13$^\textrm{th}$ to 19$^\textrm{th}$ harmonics are shown in Fig.~\ref{fig:divergenceExperiment}~(a), and compared with the predictions of the Gaussian model in (b), on the same vertical and horizontal scales. The harmonic widths in (b) were calculated as $(z_i+L)\theta_i$, where $L=6$~m is the distance from the gas cell to the measurement point. A laser waist of 450 $\mu$m and an intensity of $3.5 \times 10^{14}$ W~cm$^{-2}$ were assumed. 
The general trends observed in the experiment are reproduced by the calculations. The harmonic width generally increases as the generation position moves towards (and beyond) the laser focus along the propagation direction, and this increase generally becomes steeper with harmonic order. Differences between the experiments and the predictions of the Gaussian model could be attributed to propagation/ionization effects [see Fig.~\ref{fig:prop_tdse}~(b)], non-Gaussian fundamental beam, etc.

\section*{Chromatic aberrations of attosecond pulses}

\begin{figure}[h] 
\centering
\includegraphics[width=0.9\linewidth]{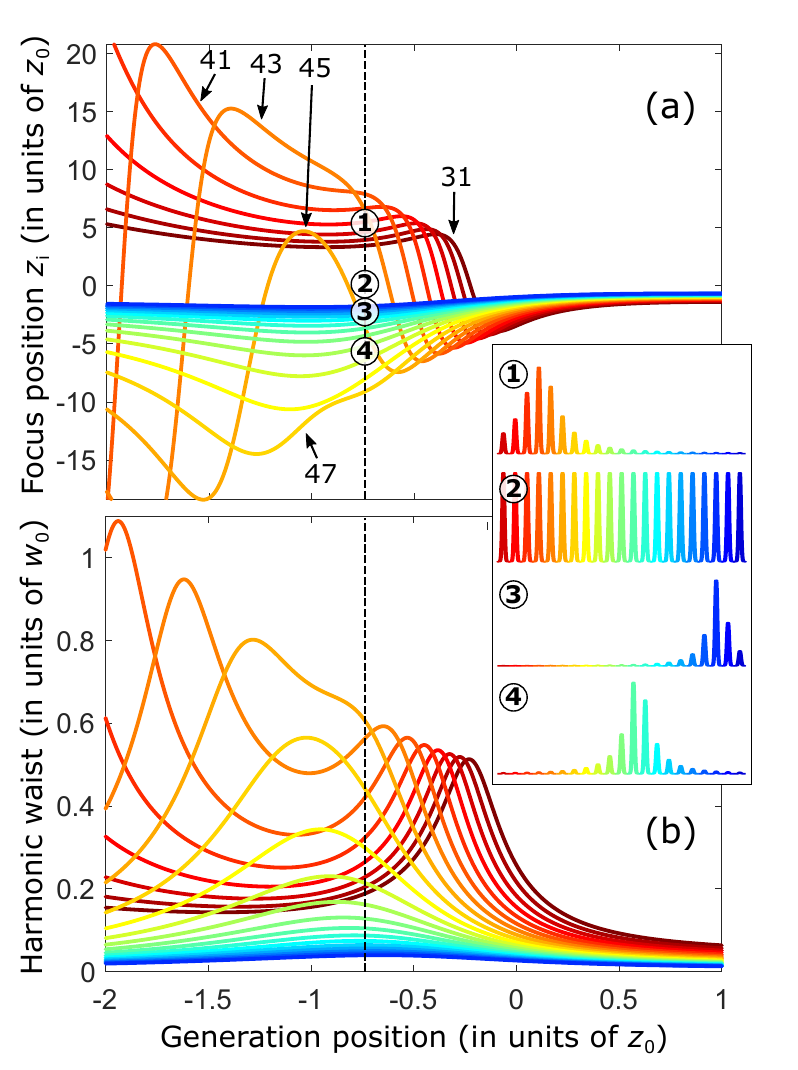}
\caption{Position of harmonic focus $z_i$ (a) and waist (b) as a function of generation position for harmonics 31 to 71. The different harmonic orders are indicated by different rainbow color codes, from brown (31) to dark blue (71). The inserts show harmonic spectra at four different positions along $z_i$, indicated from the top to the bottom by the numbered circles, for the generation position marked by the dashed line in (a).}
\label{fig:focus_harm}
\end{figure}

Finally, we study the variation of the focus position and beam waist over a large spectral bandwidth. To obtain a broad spectral region, we consider generation of high-order harmonics in neon atoms. HHG spectra obtained in Ne \cite{MacklinPRL1993} are broader and flatter than those in Ar, where a strong variation due to a Cooper minimum is observed around 45 eV.  Fig.~\ref{fig:focus_harm} shows the predictions of the Gaussian model for the 31\textsuperscript{st} to the 71\textsuperscript{st} harmonics of 800~nm radiation, at an intensity of 5 $\times 10^{14}$ W$\,$cm$^{-2}$. We only consider here the contribution from the short trajectory.  

The variation of the focus position as a function of generation position strongly depends on the process order. This is due to the frequency dependence of Eq.~[\ref{eq:radius}], and in particular depends on whether the radius of curvature diverges. Since $\mu_\mathrm{s}$ increases with frequency, the two zeros $Z_\pm$ of Eq.~[\ref{eq:radius}] move closer to each other, as is clear in Fig.~\ref{fig:focus_harm}~(a) by comparing, e.g. harmonics 41 and 43 ($Z_\pm$ correspond to the maxima in the figure). At a certain frequency, corresponding to harmonic 45 in Fig.~\ref{fig:focus_harm}~(a), $-1/\mu_\mathrm{s}$  becomes tangent to $R(z)$ at $z=-z_0$ (see also Fig.~\ref{fig:radius}). Above this frequency, the radius of curvature does not diverge and remains negative. The harmonic focus position is then always located before the generation position. As $-1/\mu_\mathrm{s}\to 0$ when the frequency increases, the focus position becomes largely independent from the generation. In this region, the harmonics are much more focused, as shown by the blue lines in Fig.~\ref{fig:focus_harm}~(b). 

To estimate the consequence of these spatial properties on the spectral characteristics of the attosecond pulses \cite{FrumkerOE2012}, we examine the variation of the on-axis spectrum at different positions (on the vertical axis), for the generation position indicated by the dashed line. This is equivalent to examining the properties of the generated radiation after refocusing as illustrated in Fig.~\ref{fig:coupling}, as a function of a ``detection position'', in the focal region. We here assume equal strength of the generated harmonics, but account for the frequency variation in beam waist size and position, [Fig.~\ref{fig:focus_harm}]. The harmonic spectra shown in the inset are found to be strongly dependent on the ``detection position'', with, in some cases, strong bandwidth reduction and displacement of the central frequency.  

\section*{Spatio--temporal coupling of attosecond pulses}

Finally, we estimate the influence of the chromatic aberrations on the temporal properties of the attosecond pulses.
We consider a  flat spectrum between harmonics 31 and 71 at the generation position indicated by the dashed line in Fig.~\ref{fig:focus_harm}.
We propagate the harmonics and coherently add them to obtain the resulting attosecond pulse train in space and time at different ``obervation'' positions. We take into account the different focus positions and divergences of the frequency components of the attosecond pulses, as well as the so-called ``attosecond'' positive chirp according to the blue curve in Fig.~\ref{fig:energy}.

\begin{figure}[h] 
\centering
\mbox{\includegraphics[width=0.95\linewidth]{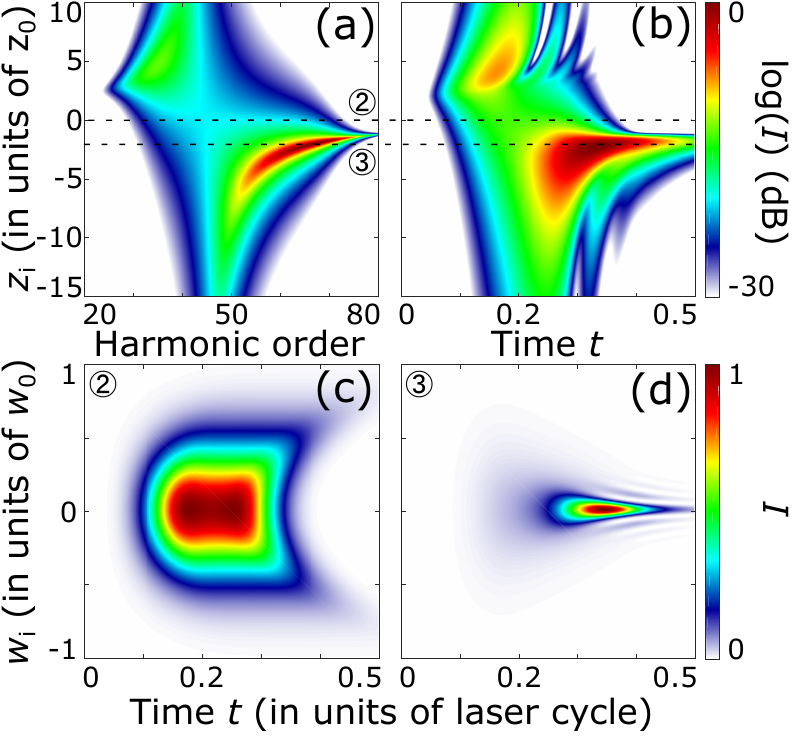}}
\caption{Graphs (a) and (b) show the on-axis spectral and temporal intensity, respectively, in logarithmic scales, as a function of the observation position, when generating at $z=-0.75$. (c) and (d) show retrieved attosecond pulses at two different observation positions, i.e. \textcircled{2} and \textcircled{3} in Fig.~\ref{fig:focus_harm}, in a linear scale.}
\label{fig:duration}
\end{figure}

Fig.~\ref{fig:duration} shows the spectral (a) and temporal (b) intensity (in color) of the generated attosecond pulse on-axis as function of the axial (observation) position relative to the generation position, here equal to $z=-0.75$ (dashed line in Fig.~\ref{fig:focus_harm}).
In these conditions, the central frequency and pulse duration of the attosecond pulse vary distinctively, indicating strong spatio--temporal couplings.
In particular, the high frequency components (high harmonic orders) form a tight virtual focus before the generation position, while the low frequency components have a more loose and real focus behind [see  Fig.~\ref{fig:duration}(a)].
The highest intensity is obtained before the generation position, while the shortest pulse is obtained afterwards, as follows from Fig.~\ref{fig:duration}(b).
The attosecond pulse is not the shortest at the generation position, where the spectral bandwidth is the largest, because the attosecond chirp stretches the pulse in time.
Fig.\ref{fig:duration}(a,b) strikingly show that the shortest pulse and the highest intensity of the attosecond pulse are obtained in different positions, illustrating the difficulty of re-focusing high-order harmonics, particularly for applications requiring high intensity.

Finally, Fig.~\ref{fig:duration}(c,d) shows the spatio-temporal intensity profiles of the attosecond pulse at the positions where is it spectrally broadest (c, \textcircled{2}) and where it is most intense (d, \textcircled{3}). The difference between the two quantities is a signature of the strong spatio-temporal couplings of the generated attosecond pulses. These couplings, here studied at  $z=-0.75$ (dashed line in Fig.~\ref{fig:focus_harm}), strongly depend on the position of generation.

\section*{Conclusion}

In this work, we examine the focusing properties of high-order harmonics generated in gases. We develop a simple Gaussian optics model based on an analytical expression of the frequency- and intensity-dependent dipole phase. This model allows us to predict the focus and divergence of the two trajectory
contributions to HHG. We validate the predictions of the model by numerical calculations based on solving the time-dependent Schr\"odinger equation for the single atom response and propagation equations for the response from the macroscopic medium. Experimental divergence measurements performed at the intense XUV beamline of the Lund Laser Centre show similar trends as those predicted by the model. We also discuss the consequences of the fact that the harmonics have different focus positions and beam waists on the resulting spectra and pulse durations. The effects investigated in the present work have a strong impact on applications of attosecond pulses, requiring a small focal spot (e.g. in order to reach a high XUV intensity) over a broad bandwidth or during a short (attosecond) duration. These spatio-temporal couplings may be reduced by locating the generation medium after the laser focus and/or by minimizing the influence of the dipole phase, using a shaped fundamental beam \cite{BoutuPRA2011} or generating in waveguides (capillaries) \cite{DurfeePRL1999,PopmintchevNP2010}.       

\section*{Funding Information}
This research was supported by the Swedish Research Council, the Swedish Foundation for Strategic Research and the European Research Council (grant 339253 PALP), the Knut and Alice Wallenberg Foundation, and the National Science Foundation (Grant No. PHY-1713761). This project received funding from the European Union's Horizon 2020 research and innovation program under Marie Skłodowska-Curie Grant Agreements no. 641789 MEDEA and 793604 ATTOPIE. 


\bibliographystyle{bibTemplate}
{\footnotesize
\bibliography{Ref_lib}
}


\end{document}